\documentclass[twocolumn,showpacs,aps,prb,superscriptaddress,floatfix]{revtex4}
\usepackage{graphicx}
\usepackage{dcolumn}
\usepackage{bm}
\begin{document}
\preprint{}
\newcommand{\scbo}{SrCu$_2$(BO$_3$)$_2$\,}
\newcommand{\dscbo}{SrCu$_{2-x}$Mg$_x$(BO$_3$)$_2$\,}
\newcommand{\scibo}{SrCu$_2$($^{11}$BO$_3$)$_2$\,}
\newcommand{\dscibo}{SrCu$_{2-x}$Mg$_x$($^{11}$BO$_3$)$_2$\,}


\affiliation{Department of Physics and Astronomy, McMaster University,
Hamilton, Ontario, L8S 4M1, Canada}
\affiliation{Canadian Institute for Advanced Research, 180 Dundas St. W.,
Toronto, Ontario, M5G 1Z8, Canada}

\author{S. Haravifard}
\affiliation{Department of Physics and Astronomy, McMaster University,
Hamilton, Ontario, L8S 4M1, Canada}

\author{S.R. Dunsiger}
\affiliation{Department of Physics and Astronomy, McMaster University,
Hamilton, Ontario, L8S 4M1, Canada}

\author{S. \surname{El Shawish}}
\affiliation{J. Stefan Institute, SI-1000 Ljubljana, Slovenia}

\author{B.D. Gaulin}
\affiliation{Department of Physics and Astronomy, McMaster University,
Hamilton, Ontario, L8S 4M1, Canada}
\affiliation{Canadian Institute for Advanced Research, 180 Dundas St. W.,
Toronto, Ontario, M5G 1Z8, Canada}

\author{H.A. Dabkowska}
\affiliation{Department of Physics and Astronomy, McMaster University,
Hamilton, Ontario, L8S 4M1, Canada}

\author{M.T.F. Telling}
\affiliation{Rutherford Appleton Laboratory, ISIS Pulsed Neutron
Facility,
Chilton, Didcot, Oxon OX110QX, UK}

\author{J. \surname{Bon\v ca}}
\affiliation{J. Stefan Institute, SI-1000 Ljubljana, Slovenia}
\affiliation{Faculty of Mathematics and Physics, University of Ljubljana, SI-1000 Ljubljana, Slovenia}


\title{In-Gap Spin Excitations and Finite Triplet Lifetimes in the Dilute
Singlet Ground State System \dscbo}

\begin{abstract}
High resolution neutron scattering measurements on a single crystal
of \dscbo with x$\sim$0.05 reveal the presence of new spin excitations within the gap of
this quasi-two dimensional, singlet ground state system.  Application of a magnetic field
induces Zeeman-split states associated with S=1/2 unpaired spins which are antiferromagnetically
correlated with the bulk singlet.   Substantial broadening of both the one and two-triplet excitations in the
doped single crystal is observed, as compared with pure \scbo. Theoretical calculations using a variational algorithm
and a single quenched magnetic vacancy on an infinite lattice are shown to qualitatively account for these effects.
\end{abstract}

\maketitle
Quasi-two dimensional quantum magnets which display collective
singlet or spin gap behavior are very topical due to the novelty of
their ground states\cite{Review} and their relation to high
temperature superconductivity in the copper oxides.  There are
relatively few such materials, and crystal growth difficulties have
further limited their study in single crystal form. \scbo is
established as a realization of the two dimensional Shastry
Sutherland model\cite{ShastrySutherland} for interacting S=1/2
dimers\cite{Miyahara, Kageyama1999}. It is comprised of well
separated layers of Cu$^{2+}$, S=1/2 orthogonal dimers arranged on a
square lattice. The material crystallizes into the tetragonal space
group I42{\it m} with room temperature lattice parameters of a=8.995
{\AA}, c=6.649 {\AA} \cite{Smith}.

\scbo  has been well studied by an array of experimental techniques, which show it to possess a non-magnetic
ground state.  In particular earlier neutron\cite{Kageyama2000, Cepas, Kakurai, Gaulin} and ESR
spectroscopy\cite{Zorko, Nojiri}
have established the leading terms in its microscopic
Hamiltonian:

\begin{eqnarray}
\mathcal{H} = J\sum_{nn} {\bf S_{i}\cdot S_{j}} + J'\sum_{nnn}{{\bf S_{i}\cdot S_{j}} + g\mu_B {\bf H} \cdot \sum_i {\bf S_{i}} }
\end{eqnarray}

\noindent
where J is the exchange interaction within the dimers and J$^\prime$ is the exchange interaction between S=1/2 spins on
neighboring dimers.  Subleading Dzyaloshinskii-Moriya interactions
weakly split the three triplet modes
even in zero applied magnetic field\cite{Cepas, Kakurai, Gaulin, ElShawish1}.

Both the exchange interactions are antiferromagnetic and their ratio x=${J^{\prime}\over J}$
has been estimated between 0.68 and 0.60, with more recent refinements being smaller.  Theoretically, such a
quantum magnet is known to possess a singlet ground state so long as the ratio, x, of inter to intra-dimer
antiferromagnetic exchange is sufficiently small\cite{Theoryreview}.  All of these estimates place \scbo
on the low side of the critical value of x at which a quantum phase transition occurs between a four sublattice
Neel state and a collective singlet state.

In finite magnetic field, much
interest has focused on a finite magnetization which develops at
fields beyond $\sim$ 20 T, wherein the lowest energy of the three
triplet states has been driven to zero energy\cite{Kageyama1999,Kodama,Onizuka,Jorge}.  The magnetic field
acts as a chemical potential for the triplet density within the
quasi-two-dimensional planes.  Magnetization plateaus ensue at
higher fields, corresponding to Bose condensation of the triplets
at certain densities.

While pure \scbo  has been well studied, there is little information
on this quantum magnet in the presence of dopants, and no reports on
doped single crystals.  This problem is very interesting by analogy
with the remarkable properties of doped quasi-two dimensional Mott
insulators and high temperature superconductivity\cite{HighTC}.
\scbo  is itself a Mott insulator and the theory of doped Mott
insulators on the Shastry-Sutherland lattice shows the possibility
of several different superconducting phases as a function of
doping\cite{ShastryKumar, Kimura, ChungKim}. Several doping studies
of \scbo  have been reported on polycrystalline samples wherein
Al, La, Na, and Y substitute at the Sr site and Mg substitutes at the Cu
site\cite{Liu}.

In this Letter we report high resolution time-of-flight neutron scattering measurements on large single crystals of
\dscibo  and \scibo.  These measurements show that doping of the magnetic Cu$^{2+}$ site with non-magnetic, isoelectronic
Mg$^{2+}$ at the 2.5 $\%$ level introduces new magnetic excitations into the singlet energy gap, and gives a finite
lifetime to all three single triplet excitations, while also substantially broadening the two triplet bound state.

\begin{figure}
\centering \resizebox{3.35in}{3.15in}{\includegraphics{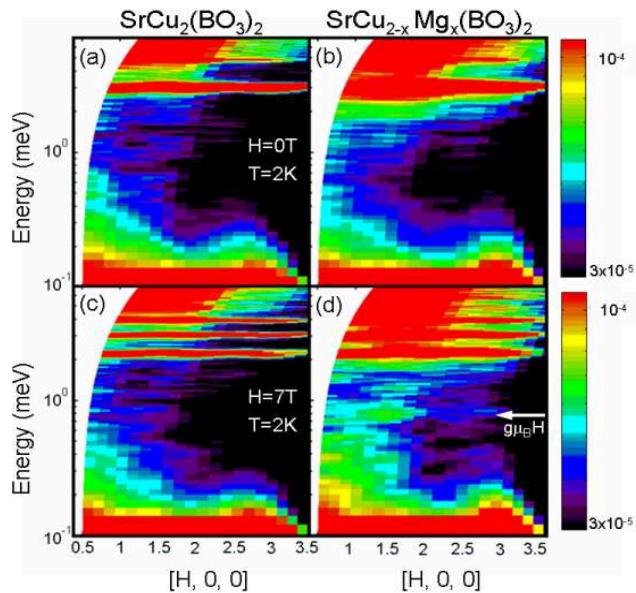}}
\caption{(a) The two left hand panels show neutron scattering data
from \scbo at T=2K.  The scattering has been integrated along L and
we show data in a magnetic field of zero and 7 T. The right hand
panels show the same data for \dscbo.  The energy axis is on a
logarithmic scale. The Zeeman-split S=1/2 level in \dscbo in H=7 T
at g$\mu_B$H=0.8 meV is indicated in panel d).} \label{Figure 1}
\end{figure}

Two single crystal samples, \scibo and  \dscibo, were grown by floating zone image
furnace techniques at a rate of 0.2 mm/hour in an O$_2$ atmosphere.  The crystals
were of almost identical cylindrical shape, with approximate dimensions of 4.5
cm in length by 0.6 cm in diameter.  These samples were grown using $^{11}B$, to avoid the
high neutron absorption cross section of natural boron.  The pure \scibo single crystal was the same
high quality single crystal studied previously\cite{Gaulin}. Time-of-flight neutron scattering measurements were
performed using the OSIRIS spectrometer\cite{Telling} at the ISIS Pulsed
Neutron Source of the Rutherford Appleton Laboratory.  OSIRIS is an
indirect geometry time-of-flight spectrometer which employs an array of pyrolytic
graphite monochromators to energy analyse the scattered neutron beam.  The data
was collected with the spectrometer configured to use the 004 analysing reflection afforded by
pyrolytic graphite.  In this configuration, only those neutrons with a scattered energy of 7.375 meV
are Bragg reflected towards the detector.
The [H,0,L] plane of both crystals was coincident with the horizontal scattering plane, and the
samples were mounted in a 7 T vertical, [0,K,0], field magnet cryostat.

Figure 1 shows representative time-of-flight neutron scattering
data, taken at T=2 K and H=0 and 7 T for \scbo (a and c) and for \dscbo (b and d).
This data was integrated
along L, in which direction the spin excitations show little
dispersion\cite{Gaulin}. Note the logarithmic energy and intensity
scales, chosen to draw attention to the
details of the in-gap excitations seen in \dscbo.  The splitting of
the triplet excitations near 3 meV on application of the 7 T
magnetic field is clear.  In finite field, weak
dispersion of the triplet excitations as a function of wavevector H
is seen, and this has been attributed to subleading terms in the
spin Hamiltonian - terms other than those in Eq. 1. There are
several qualitative features evident on examination of this data.
The one triplet excitations show significantly greater breadth in
energy in \dscbo than in \scbo.  In addition, application of a H=7 T
magnetic field gives rise to an inelastic peak at
g$\mu_B$H =0.8 meV in H=7 T, which is centered around [H$\sim$ 1.4,
0, 0] in {\bf Q}-space, but extends in wavevector H to almost
[H$\sim$3, 0, 0]. Both of these features are discussed at length
below.

\begin{figure}
\centering \resizebox{3.35in}{2.78in}{\includegraphics{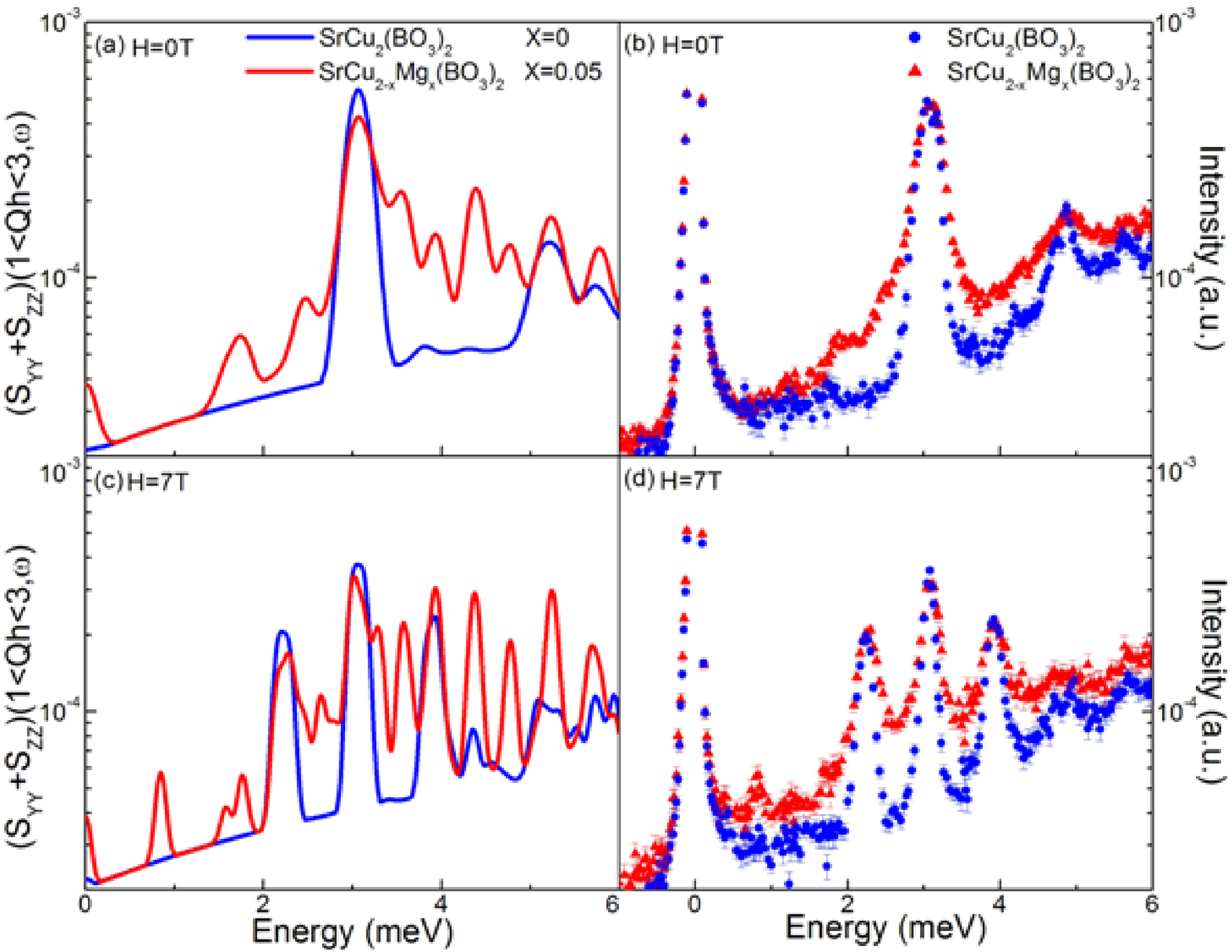}}
\caption{The right hand panels show inelastic neutron scattering
data for both \scbo and \dscbo in H=0 (b) and H=7 T (d). This data
has been integrated in {\bf Q} between H=1 and H=3 and over all L.
The intensity is plotted on a logarithmic scale. Clearly the one and
two triplet excitations are considerably broader in energy in \dscbo
as compared with \scbo, and in-gap states are introduced on doping.
The left hand panels show the numerical calculation for S$^{zz}({\bf
Q}, \omega)$ + S$^{yy}({\bf Q}, \omega)$, using J=76.8 K and
J$^\prime$/J=0.62, which accounts for many of the qualitative
features observed in \dscbo.} \label{Figure 2}
\end{figure}

Figures 2b and 2d show the same experimental data
as in Fig. 1, now integrated in wavevector along L and also in H
between H=1 rlu and H=3 rlu, and plotted as a function of energy.  The
data in a magnetic field of 0 and 7T is shown in Figs 2b and 2d, respectively.
The extra breadth in both the single triplet excitations
and the two triplet bound states above it is clear. Broad inelastic peaks appear within the gap, which possess
little magnetic field dependence, indicating a longitudinal nature.  Sharp,
field-induced inelastic scattering at $\hbar\omega \sim$ 0.8 meV is also evident.

Numerical calculations have also been carried out using a new
variational approach\cite{ElShawish} to solve the model of a single quenched
impurity on the two dimensional Shastry-Sutherland lattice.  This
method generates a variational space by successively applying the
off-diagonal parts of the Hamiltonian, Eq. 1, on the starting
approximation for the single impurity ground state, which consists of a
free spin 1/2 neighboring the impurity site, embedded within a
dimer background. The resulting small spin polaron structure and
exponential growth of the variational space with each iteration
guarantee good convergence of the spin polaron ground state, as well
as for the lowest energy excited states.  Full details will be given
separately\cite{ElShawish}.  For energies below $\sim$ 3 meV
the method provides accurate and converged results for both the longitudinal,
S$^{zz}({\bf Q}, \omega)$, as well as the transverse, S$^{yy}({\bf
Q}, \omega)$ components of the dynamical spin structure factor.
These are compared directly to the neutron scattering
experiments in Fig. 2.

Figures 2a and 2c show the calculated S$^{zz}({\bf Q}, \omega)$ +
S$^{yy}({\bf Q}, \omega)$, integrated over the same range of
wavevectors as the experimental data, and at magnetic fields of 0 (a) and 7 T (c).
The comparison between theory and experiment is qualitatively good.  The numerical results confirm
that quenched magnetic vacancies induce in-gap states, substantial spectral weight
below the zero field gap energy of $\sim$ 3 meV.  A component of these in-gap states show
little magnetic field dependence, and appear in the S$^{zz}({\bf Q}, \omega)$, longitudinal channel.
The calculation also shows the g$\mu_B$H transverse, S$^{yy}({\bf Q}, \omega)$, excitation in finite magnetic field.
Furthermore, the calculation\cite{ElShawish} allows a determination of the spatial distribution of the spin polaron
S=1/2 degree of freedom and its low lying excited states, and these can guide the interpretation of the
{\bf Q}-dependence of the field-induced g$\mu_B$H transverse spin excitation observed in the experiments.

\begin{figure}
\centering \resizebox{3.35in}{2.4in}{\includegraphics{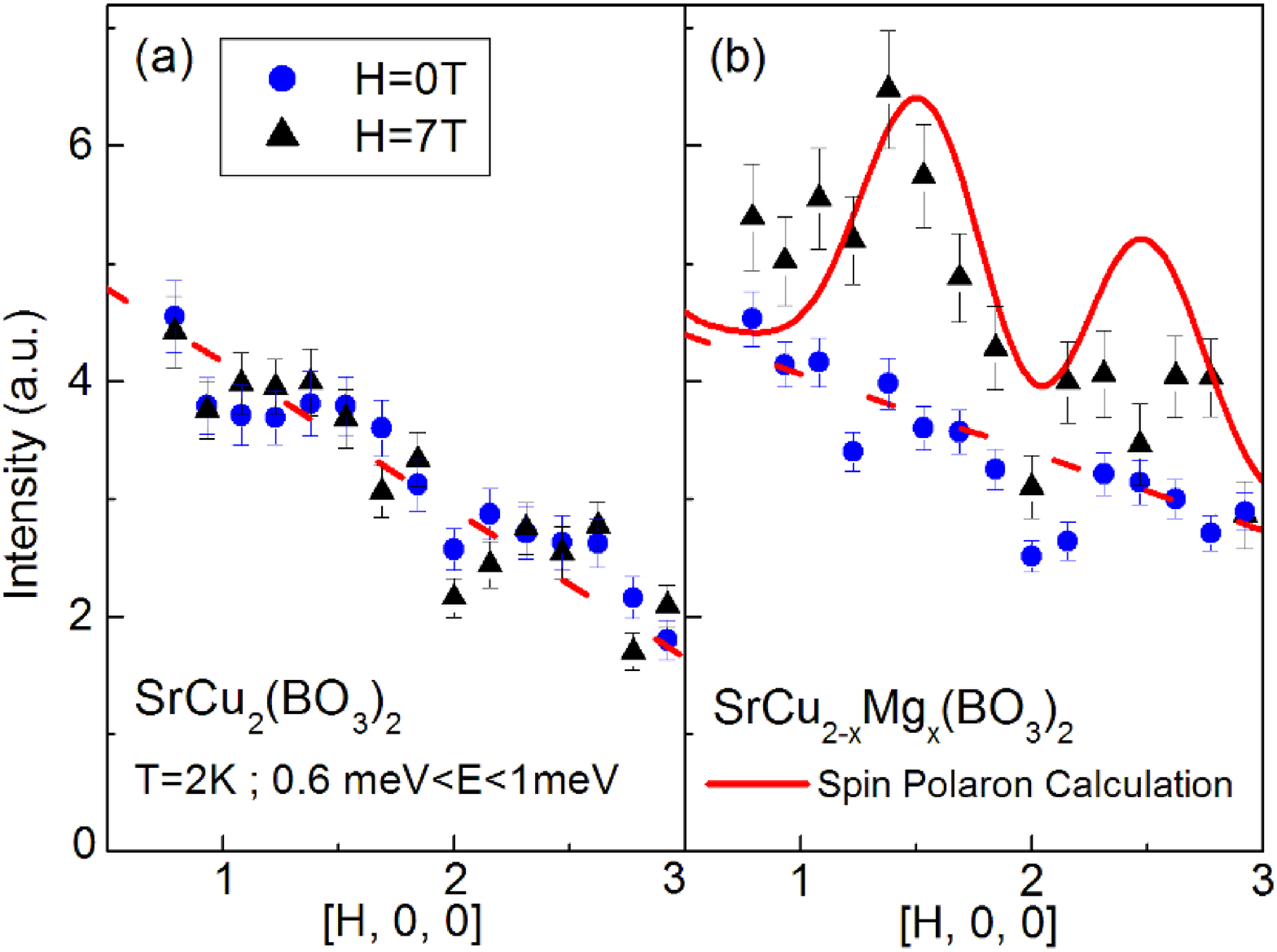}}
\caption{{\bf Q}-scans are shown at T=2 K for both \scbo (a) and
\dscbo (b) which integrate in energy between 0.6 meV and 1 meV.
This energy range captures the Zeeman energy g$\mu_B$H=0.8 meV
appropriate to g=2 and H=7 T.  (b) shows a comparison between the
experimental {\bf Q}-dependence of this scattering with the
calculated form described in the text. The field-induced in-gap
state in \dscbo is absent in \scbo.} \label{Figure 3}
\end{figure}

Figure 3 shows the {\bf Q} dependence of the scattering around
g$\mu_B$H=0.8 meV in an applied magnetic field of 7 T in both
\scbo (a) and \dscbo (b).  This scattering is
integrated in wavevector over L, and in energy between 0.6
meV $<$ $\hbar\omega$ $<$ 1 meV and is shown in a magnetic
field of both 0 and 7T. Figure 3a shows the absence of a magnetic field induced signal
in \scbo within this
energy range.  Figure 3b shows a clear field induced signal in \dscbo
which peaks at [H$\sim$1.4, 0,
0], but extends out to almost [H$\sim$3.0, 0, 0]. This field induced
scattering is attributed to Zeeman split S=1/2 states associated
with the S=1/2 moment in a dimer whose
partner site is occupied by a quenched,
nonmagnetic Mg$^{2+}$ ion.  This field induced inelastic scattering is very similar to that
associated with end states in Haldane spin chains, such as occur in
Y$_2$BaNi$_{1-x}$Mg$_x$O$_5$\cite{Kenzelmann}.  In this case,
quenched, non-magnetic Mg$^{2+}$ ions produce finite S=1 magnetic
chains.  Spin 1/2 degrees of freedom arise at the end of finite chains
of S=1 magnetic moments, as one of the two effective S=1/2 degrees
of freedom making up the S=1 moments lacks a partner with which to
form a singlet.  Such excitations occur at an energy of g$\mu_B$H in
finite field, and display a wavevector dependence which indicates
antiferromagnetic correlations into the collective singlet of the chain
segment.

The wavevector dependence  of the magnetic field-induced spin excitation at 0.8 meV can be attributed to
the structure of the spin polaron\cite{ElShawish}, whose ground state possesses strong antiferromagnetic correlations
with neighboring dimers, transverse
to the dimer containing the impurity site.  To first non-trivial order in $J'/J$, we obtain an analytical expression for
the square of the matrix element for this transition:
\begin{eqnarray}
 I^{\pm}\propto
 \Big|(a_3^2 - a_1^2/2)\,e^{{\rm i}\eta({\rm H}\pm{\rm K})} + 2\sqrt{2}\ a_1 a_2 \times\nonumber\\
 \sin(\eta({\rm H}\mp{\rm K}))\sin(\pi ({\rm H}\mp{\rm K}))
 -2\sqrt{2}\ a_2 a_3 \times\nonumber\\
 \cos(\eta ({\rm H}\mp{\rm K})) \cos(\pi ({\rm H}\mp{\rm K})) \Big|^2
  \label{qdisp}
\end{eqnarray}
where $a_1$, $a_2$, and $a_3$ are the weights of the polaron variational wavefunction computed for $J'/J=0.62$\cite{avalues}.
$\eta=0.72$
accounts for microscopic distances in SrCu$_2$(BO$_3$)$_2$. The $\pm$ sign in Eq.~(\ref{qdisp})
distinguishes between the two nonequivalent impurity positions
within the unit cell. Random doping therefore implies $I\propto I^{+}+I^{-}$, which we show, multiplied by the
square of the magnetic form factor for Cu$^{2+}$, in Fig. 3b
along with measurements on {\bf Q}-dependence of this scattering in SrCu$_{2-x}$Mg$_x$(BO$_3$)$_2$.  While the agreement between
the calculation and experiment is not perfect, the calculation captures the general {\bf Q}-dependence of the excitation.

\begin{figure}
\resizebox{3.4in}{2.85in}{\includegraphics{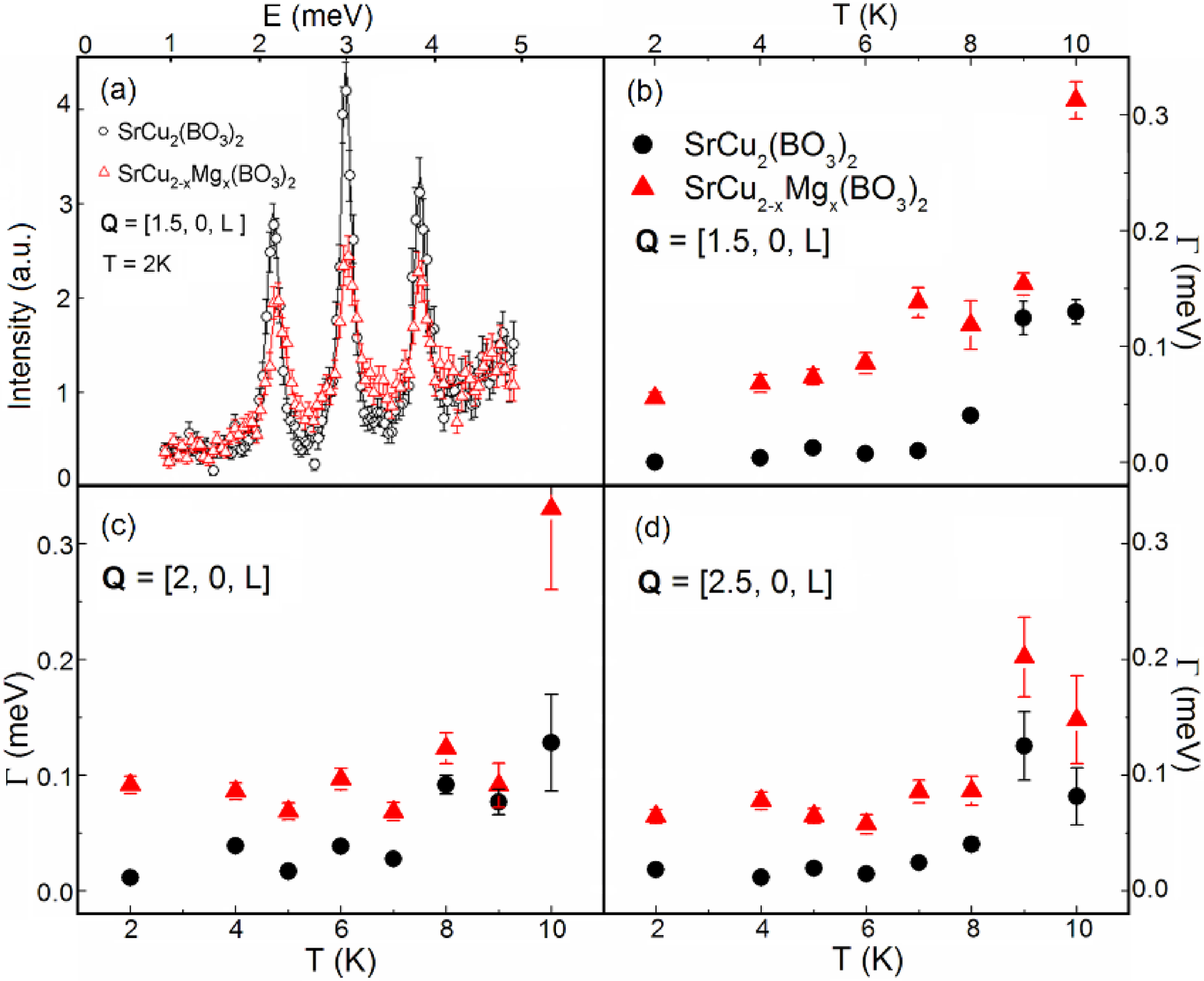}} \caption{(a)
Cuts of the data simulating constant-{\bf Q} scans at (1.5,0,L),
integrated along L, are shown for \scbo and \dscbo. Resolution
convoluted fits to the data are shown as the solid lines and the
description of the data is excellent. Such fits allow us to extract
the triplet excitation lifetimes ($\Gamma$), and which are shown in
(b), (c), and d) for wavevectors (1.5, 0, L), (2, 0, L) and (2.5, 0,
L) respectively, as a function of temperature. Even at the lowest
temperatures, we observe finite triplet lifetimes in \dscbo.}
\label{Figure 4.}
\end{figure}

Figure 4a shows representative data
with accompanying fits used to extract the lifetimes of
the one triplet excitations as a function of doping and temperature in a magnetic field of 7 T.
This data is integrated in {\bf Q} along L, and over a narrow range in wavevector H around H=1.5, and these cuts
approximate constant-{\bf Q} scans.  Similar analysis was performed around wavevector H=2.0 and 2.5 to look for
variation in the one-triplet lifetimes as a function of {\bf Q} along [H, 0, 0].  The data
was fit to the sum of three identical damped harmonic oscillators (DHO):

\begin{eqnarray}
S(\textbf{Q},\omega,T) = \chi(\textbf{Q},T)\frac{1}{1-exp\left(-\frac{\omega}{kT}\right)}\times\nonumber\\
\left[\frac{4\omega\Gamma_{\textbf{Q},T}/{\pi}}{\left(\omega^{2}-\Omega^{2}_{{\textbf{Q},T}}\right)^{2} + 4\omega^{2}\Gamma^{2}_{\textbf{Q},T}}\right]
\end{eqnarray}

\noindent
where $\chi$(\textbf{Q}, \textit{T}) is the momentum-dependent
susceptibility, while the second term takes into account detailed
balance.  The renormalized DHO frequency, $\Omega_{{\bf Q}}$, has contributions from the oscillation
frequency, $\omega_{{\bf Q},T}$, and the damping coefficient, $\Gamma_{{\bf Q},T}$, and is given by:
$\Omega^{2}_{\textbf{Q},T} = \omega^{2}_{\textbf{Q},T} + \Gamma^{2}_{\textbf{Q},T}$.

Equation 3 was convoluted with an appropriate resolution function and fit to the data.
The results are shown in Figs. 4b), c), and d), for wavevectors H=1.5, 2.0, and 2.5
respectively. The
extracted lifetimes show little or no systematic variation with
wavevector H, but a finite one-triplet lifetime is observed in
\dscbo at all temperatures, in contrast to \scbo, where the
low temperature one-triplet lifetimes are very long, compared with
the resolution of the spectrometer.  In both \dscbo and \scbo, the
thermal destruction of the collective singlet ground state near
$\sim$ 10 K is characterized by a rapid decrease in the one triplet
lifetime (1/$\Gamma$) on warming, with little or no softening of
the one triplet excitation energies.

As mentioned previously, the theoretical results for S$^{zz}({\bf Q}, \omega)$ + S$^{yy}({\bf Q}, \omega)$
are not well converged for energies of $\sim$ 3 meV and greater.  Nonetheless, the additional broad spectral weight around
the calculated one and two triplet energies seen in Fig 2a) and c) is consistent with finite
triplet lifetimes in the presence of a single quenched magnetic vacancy.

To conclude, new inelastic neutron scattering measurements on \dscbo with x$\sim$ 0.05 show
relatively broad and field-independent in-gap spin excitations as well as a magnetic field-induced excitation
identified as a Zeeman-split, spin polaron state.  The non-magnetic quenched vacancies also give
rise to finite and measurable lifetimes in the one and two triplet excitations.

We wish to acknowledge very helpful discussions with T.G. Perring and expert technical support
from the ISIS User Group. This work was supported by NSERC of Canada, and the Slovenian Research Agency
under contract PI-0044.
%
%
%
%
%
%
%
%
%
%

\end{document}